\documentclass[fleqn,twoside,twocolumn,nofootinbib,showkeys]{revtex4}
\usepackage[sec,nocpr]{ujp} 

\begin{document}
\title[Common Approaches in Description of Ordinary Liquids]
{COMMON APPROACHES IN DESCRIPTION\\ OF ORDINARY LIQUIDS AND HADRONIC
MATTER}
\author{K.V.~Cherevko}
\affiliation{Faculty of Physics, Taras Shevchenko National University of Kyiv}
\address{4, Academician Glushkov Ave., Kyiv 03022, Ukraine}
\email{k.cherevko@univ.kiev.ua}
\author{L.L.~Jenkovszky}
\affiliation{Bogolyubov Institute for Theoretical Physics, Nat.
Acad. of Sci. of Ukraine}
\address{Metrolohichna Str. 14-b, Kyiv, 03680, Ukraine}
\email{jenk@bitp.kiev.ua}
\author{V.M.~Sysoev}
\affiliation{Faculty of Physics, Taras Shevchenko National University of Kyiv}
\address{4, Academician Glushkov Ave., Kyiv 03022, Ukraine}
\email{k.cherevko@univ.kiev.ua}
\author{Feng-Shou Zhang\,}
\affiliation{College of Nuclear Science and Technology, Beijing
Normal University }
\address{Beijing 100875, China}
\affiliation{Beijing Radiation Center}
\address{Beijing 100875, China}
\affiliation{Center of Theoretical Nuclear Physics, National Laboratory of Heavy Ion Accelerator of Lanzhou}
\address{Lanzhou 730000, China}
\email{fszhang@bnu.edu.cn}

\udk{536.7,539.1,524.8} \pacs{05.70.-a, 21.65.-f,\\[-3pt] 25.70.-z,
25.75.-q} \razd{\secii}

\autorcol{K.V.\hspace*{0.7mm}Cherevko,
L.L.\hspace*{0.7mm}Jenkovszky, V.M.\hspace*{0.7mm}Sysoev et al.}

\setcounter{page}{708}%

\begin{abstract}
This work is an attempt to give a brief overview of the
implementation of the statistical thermodynamics to hadronic matter.
The possibility to use the hydrodynamic approach for developing the
physical model of the formation of exotic structures in the head-on
intermediate-energy heavy ion collisions is discussed.\,\,That
approach is shown to be able to provide simple analytical
expressions describing each step of the collision process.\,\,This
allows for extracting the data concerning nuclear matter properties
(surface tension, compressibility, {\textit{etc.}}) from the
properties of  observed fragments.\,\,The advantages of the
thermodynamic analysis of phase trajectories of the system in heavy
ion collisions are discussed.\,\,Within the thermodynamic approach,
the method to evaluate the curvature correction to the surface
tension from the nuclear matter equation of state is
described.\,\,The possibility to use statistical thermodynamics in
the studies of hadronic matter and quantum liquids is discussed.
\end{abstract}


\maketitle

\section{Introduction}

The study  of the properties of hadronic matter and the phase
transitions in it has been the subject of many studies both
theoretical and experimental in recent decades.\,\,Such an interest
is explained both by the applied importance of results for the
energy production industry, radiation medicine and by their
fundamental importance, because the many-body systems, in which the
excitation energy is thermally distributed among many particles and
in which there exist long-range Coulomb and short-range forces, up
to now have no satisfactory description.\,\,The description of
nuclear matter (NM) within statistical thermodynamics adopted in the
recent decades is very similar to that for ordinary liquids
\cite{Bulavin2010,Bulavin2006,Bulavin2005,Gross2001}.\,\,The
physical reason for that is the similarity between the molecular
forces and nuclear-nuclear interactions in respect to their
dependence on the distance.\,\,Notwithstanding the tremendous
difference in the energy and space scales, the nuclear and van der
Waals forces represent a very similar behavior.\,\,In both cases,
the attraction between particles is replaced by their repulsion at a
small interaction range.\,\,J.D.\,\,van der Waals suggested his
famous equation in 1873.\,\,A hundred years later, his finding was
fruitfully used to describe the properties of the nuclear matter
unknown to the nineteenth century scientists.\,\,It is quite natural
to expect that, when investigating both ordinary liquids and nuclear
matter, it is reasonable to use similar fundamental concepts of the
modern thermodynamics, statistical physics, phase transitions
theory, and the liquid state theory accounting for the quantum
nature of nuclear matter.

Among all the studies  of nuclear systems, of high importance are
those devoted to the processes observed in heavy ion collisions
(HIC).\,\,Such studies are valuable from the fundamental point of
view, as it is the only available experimental method (under the
Earth conditions) to obtain data about the nuclear matter at extreme
temperatures and densities.\,\,As for the applied importance, the
results are needed to proceed in developing the new technologies of
the nuclear waste utilization, which is an up-to-date task
accounting for the rapid development of the nuclear energy
production industry.\,\,The other important applications can be
found in medical physics and biophysics.\,\,When the beams of
high-energy ions pass through the matter, they leave the trace of
the fragments produced in the collision, causing the changes in
structures and thermodynamic properties of biological objects and
liquids.\,\,Therefore, the clear understanding of the fragments
formation mechanisms can help in biophysical studies and in
developing the new technologies in the medical physics.

Nowadays, it could be seen that, in the field of studies devoted to
the HIC, the progress in experimental researches is much faster than
in theoretical ones.\,\,Unfortunately, there are yet no complete
theories of such important phenomenon as the nuclear
multifragmentation, no satisfactory equation of state (EOS) for
hadronic matter, and no clear understanding of phase transitions in
nuclear matter.\,\,Therefore, the progress in the theory able to
describe the basic thermodynamic and structural properties of
hadronic matter, phase transitions in it, and behavior of hot nuclei
with high excitation energies stands among the most important
problems in modern physics.

To recall the stage, we would  like to present a brief overview of
the problems that exist in hadronic matter physics and can be
studied with the help of statistical thermodynamics and the theory of phase
transitions well developed for ordinary liquids.

\subsection{Nuclear matter equation of state}

\begin{figure}[b]
\vskip-2mm
\includegraphics[width=7.3cm]{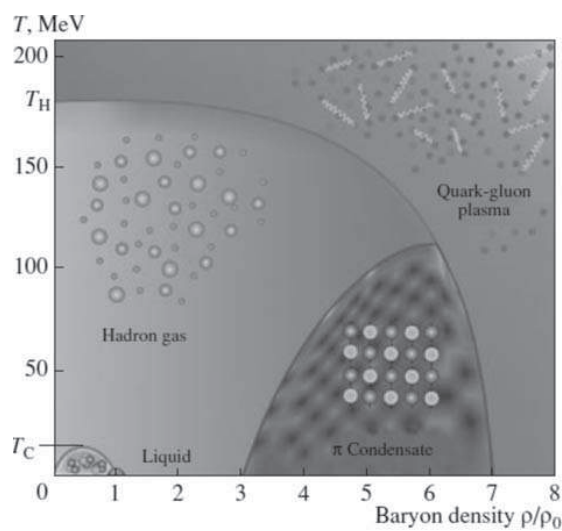}
\vskip-3mm\caption{Phase diagram of NM \cite{Karnaukhov2006}}
\label{Phase}\vskip1.5mm
\end{figure}
The knowledge of the EOS of nuclear matter is one of the fundamental
goals in nuclear physics, which has not yet been achieved
\cite{Bertsch1988, Baldo2007, Abe1996, Baran2005, Li2008a}.\,\,It
should describe the fundamental properties of NM
\begin{equation}
P=P(\rho, T, \delta),
\label{eq1b}
\end{equation}
where $P$ is the pressure, $\rho$ is the density, and $\delta$ is an
asymmetry parameter, at any point of the phase diagram of NM
(Fig.~\ref{Phase}).\,\,The EOS of NM is an important ingredient in
studying the properties of nuclei at and far from stability,
studying the structures and the evolution of compact astrophysical
objects such as neutron stars and core-collapse supernovae, applied
nuclear physics, \textit{etc.}\,\,The saturation point of EOS, i.e.,
the energy as a function of the matter density, for the symmetric NM
(SNM) at zero temperature ($T$ = 0), is well determined from the
ground-state properties of nuclei, such as binding energies and
central matter densities, by extrapolation to infinite NM
\cite{Shlomo2012}.\,\,At the same time, one can find the vast
majority of different parametrizations and different theoretical
models that were built in attempt to describe the NM and finite
nuclei in a wide range of the external parameters.\,\,All of them
were built for the case of special assumptions and are lacking a
predictive power \cite{Stevenson2012}.

\begin{figure*}
\vskip1mm
\includegraphics[width=16.7cm]{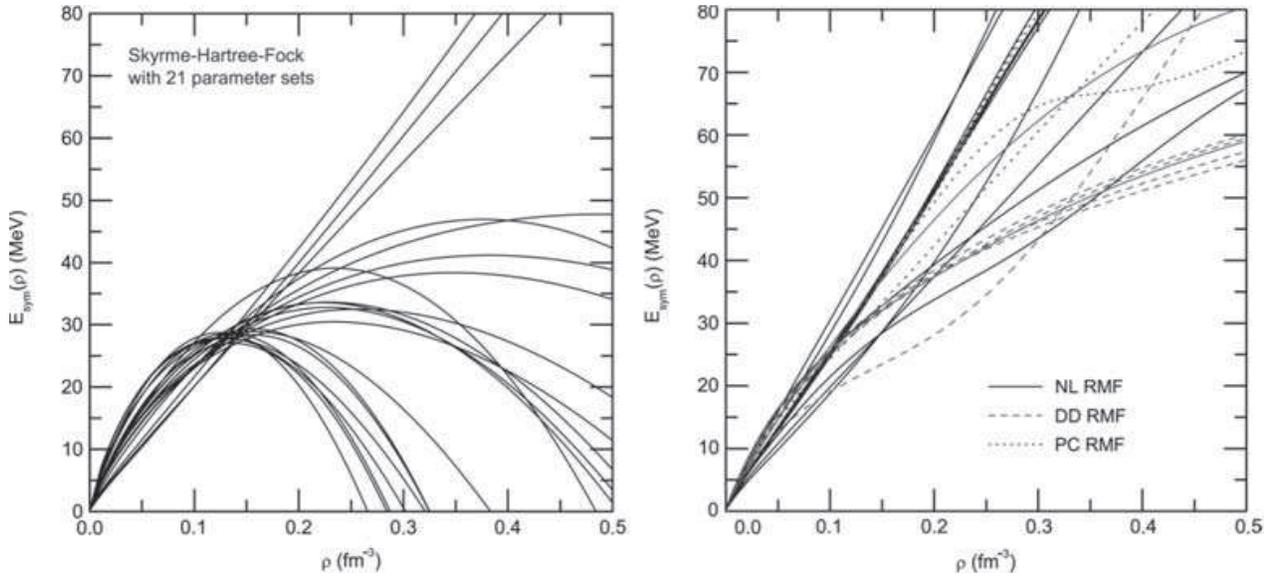}
\vskip-3mm \caption{Left window: Density dependence of the nuclear
symmetry energy $E_{sym}(\rho)$ from SHF with 21 sets of Skyrme
interaction parameters \cite{Chen2005}.\,\,Right window: Same as
left panel from the RMF model for the parameter sets NL1, NL2, NL3,
NL-SH, TM1, PK1, FSU-Gold, HA, NL$\rho$, and NL$\rho\delta$ in the
nonlinear RMF model (solid curves); TW99, DD-ME1, DD-ME2, PKDD, DD,
DD-F, and DDRH-corr in the density-dependent RMF model (dashed
curves); and PC-F1, PC-F2, PC-F3, PC-F4, PC-LA, and FKVW in the
point-coupling RMF model (dotted curves) \cite{Chen2007}}
\label{SymGraph1}
\end{figure*}

The things go even worth, when one changes to the asymmetric NM and
the pure neutron matter. Existing models give different results for
the symmetry energy of asymmetric NM \cite{Oyamatsu2000},  its
slope, and magnitude at the saturation density and, the experimental
constraints are quite wide \cite{Tsang2012}.\,\,It should be noted
that the symmetry energy behavior is very different for different
EOS parametrizations \cite{Li2008a} in the high- and low-density
regions (Fig.~\ref{SymGraph1}).

\begin{figure}
\vskip1mm
\includegraphics[width=7.5cm]{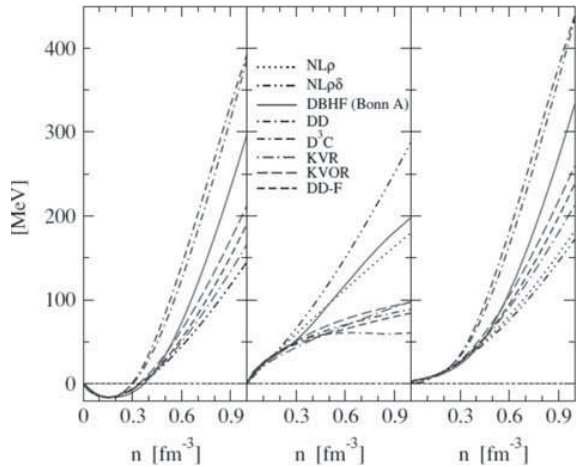}
\vskip-3mm \caption{Energy per nucleon in SNM $E_0(n)$ (left panel),
the symmetry energy $E_s (n)$ (middle panel), and the energy per
nucleon in a nonsymmetric NM (NNM) (В-equilibrated and charge
neutral) for different models (right panel) \cite{Klahn2006}}
\label{SNMGraph}
\end{figure}

The similar uncertainty is observed for a symmetric NM (SNM) in the high-density region (Fig.~\ref{SNMGraph}).

Nowadays, extracting the information on the nuc\-le\-ar EOS, in
particular at high baryon densities, is restricted to observations
of astrophysical compact ob\-jects and studies of hot nuclear
systems created in high-ener\-gy pro\-ton-in\-du\-ced reactions or
in in\-ter\-me\-dia\-te
and relativistic energy hea\-vy-ion collisions~(HIC).

Therefore,  HIC can shed light on what are nuclei made of and
explain the properties of the matter that form the nuclei.\,\,Such
experiments can be treated as a kind of materials science similar to
the $P$-$V$-$T$ studies in the physics of ordinary
liquids.\,\,Hence, it is quite natural to apply the well-developed
theories of ordinary liquids to the description of HIC.
Understanding the behavior of NM in the HIC is of great relevance
either for theoretical nuclear physics or for nuclear astrophysics,
when studying the supernova explosions and the properties of neutron
stars \cite{Botvina2009, Souza2009, Danielewicz2002}.\vspace*{-2mm}

\section{Heavy Ion Collisions}
The  idea of HIC experiments is to collide two nucleus or light
particles with nuclei at a high enough energy to break them into
pieces.\,\,Introducing then an appropriate model of the process
involved, it should be possible to extract some data on NM
properties and its behavior at high densities and
temperatures.\,\,The basic mechanisms of the formation of fragments
in the HIC experiments are the following \cite{Hufner1985}:

\raisebox{0.5mm}[0cm][0cm]{{\footnotesize$\bullet$}}\,\,spallation,
when only one heavy fragment with its mass close to the target mass
is formed ($m = 1$);

\raisebox{0.5mm}[0cm][0cm]{{\footnotesize$\bullet$}}\,\,fission that
goes usually with the two heavy fragments produced with masses in
the interval around half the target mass $A\sim\frac{A_T}{2}$;

\raisebox{0.5mm}[0cm][0cm]{{\footnotesize$\bullet$}}\,\,multifragmentation
that is the process which leads to the formation of several
fragments with intermediate masses.\,\,Usually with $A < 50$.

The last one is especially important in studies of the NM EOS.
Disintegration, when a bigger nucleus breaks into one or several
nuclei and some nucleons, is one of the two basic mechanisms in the
Nature, by which a nuclei can be formed.\,\,The appropriate way to
study this phenomenon is the experiment with proton-nucleus and
nucleus-nucleus collisions.\,\,When such a collision is observed at
a high enough energy (higher than the threshold value $E_{\rm
th}\sim$~2--4~MeV/nucleon \cite{Botvina2006}), there is a high
probability of the production of intermediate-mass fragments
\cite{Kwiatkowski1994, Pan1995}.\,\,The nature of the production of
intermediate-mass fragments in such experiments is of great interest
for the consequent understanding of the NM properties and for the
progress in the studies of nuclear EOS \cite{Pan1995, Viola2004,
Chase1995, Fuchs2006, Das2003a, Chaudhuri2009}.\,\,Thus, HIC have
been intensively studied in the last years \cite{Siemens1983,
Viola2004}.\,\,It should be noted that, in a composite system formed
in the collision process from the projectile and target nuclei,
nuclear matter can be highly compressed and be at high temperatures
during the early stage of the reaction.\,\,At the time of the
formation of intermediate-mass fragments with $3\leq{Z}\leq20$,
their characteristic properties such as the excitation energy and
isotopic distributions are governed by the characteristics of the
break-up source such as the temperature, density, and $N/Z$ ratio.
Therefore, intermediate-mass fragments may provide a unique probe to
study the reaction mechanism and hot nuclear matter properties
\cite{Rodrigues2011}.

\begin{figure}
\vskip1mm
\includegraphics[width=6.3cm]{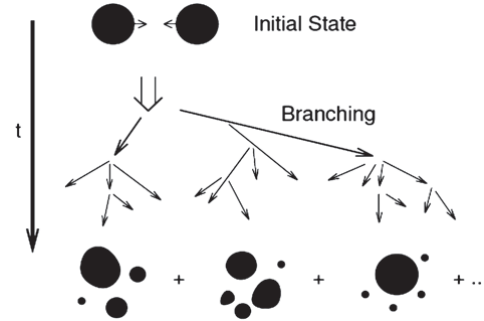}
\vskip-3mm \caption{Schematic picture of a fragmentation  reaction,
in which a given initial channel may develop into many different
fragmentation channels during the dynamical evolution \cite{Ono2006}
} \label{mf1}\vspace*{-2mm}
\end{figure}

At the same  time, in order to have the complete picture of the
situation in a field, one should bear in mind that the colliding
system is over a large time span of the reaction out of global and
even local equilibrium \cite{Fuchs2006}.\,\,That means the presence
of the non-equilibrium effects all over the compression phase, where
one essentially intends to study the NM EOS.\,\,The other important
challenge is that the heavy ion collision experiments are inclusive
and will remain so in the foreseeable future.\,\,That means missing
some information about the physical processes going on in the
system. Only the degree of inclusiveness may change if more
sophisticated counter arrays are introduced~\cite{Hufner1985}.

During the last decades, a wealth of data have been accumulated
about the multifragmentation phenomenon in proton-nucleus and
nucleus-nucleus collisions \cite{Hufner1985, Borderie2008,
Karnaukhov2006, Marie1998}.\,\,Different models (statistical
\cite{Jaqaman1991, Souza2009}, percolation \cite{Campi1986}) and
different mechanisms (liquid-gas phase transitions, spinodal
decomposition, sequential evaporation from the expanding-emitting
source, {\it etc}.\,\,\cite{Berkenbusch2001}) are being used for
explaining the existing data.\,\,As for the initial highly
non-equilibrium stage, the most common approach is using the
transport models for its adequate description.\,\,One of the
important points in that case  is allowing the model for the
possibility for the occurrence of dynamical bifurcations (Fig.
\ref{mf1}), as it is a non-linear process \cite{Ono2006}.

\begin{figure}
\vskip1mm
\includegraphics[width=7.5cm]{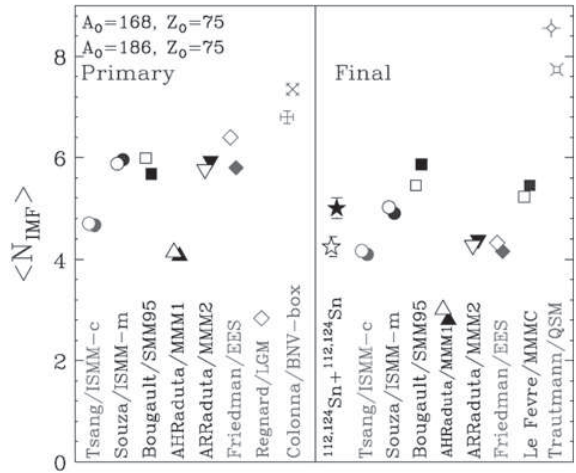}
\vskip-3mm \caption{Mean IMF multiplicity obtained from different
statistical models.\,\,The open and solid stars on the right panel
are data from the central collisions of $^{112}$Sn + $^{112}$Sn and
\mbox{$^{124}$Sn + $^{124}$Sn} systems at $E=A = 50$~MeV
\cite{Tsang2006}} \label{mf2}
\end{figure}
\begin{figure}
\vskip1mm
\includegraphics[width=7.5cm]{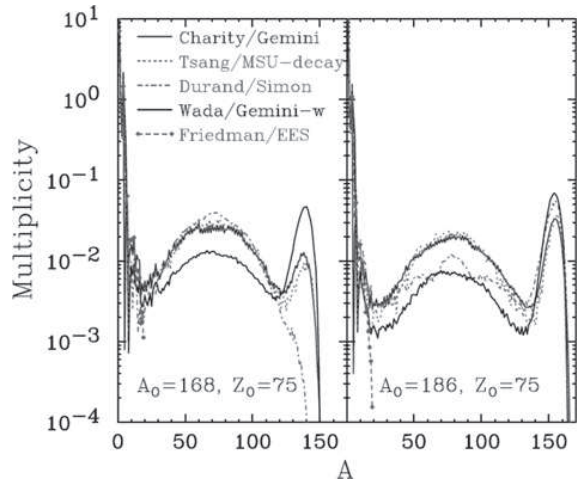}
\vskip-3mm \caption{Predicted mass distributions from five
evaporation codes for neutron-deficient system 1 (left panel) and
neutron-rich system 2 (right panel)~\cite{Tsang2006}}
\label{mf4}\vspace*{-2mm}
\end{figure}

There exist a number of different transport models that have been
developed till now.\,\,Among them are intranuclear cascade
calculations, time-dependent Hartee--Fock model, Boltzmann-type
kinetic equations based on hadronic non-equilibrium quantum
transport field theory, molecular dynamics methods (MD, QMD, AMD,
FMD), nuclear fluid dynamics calculations, {\textit{etc.}} The
detailed analysis can be found elsewhere ({{e.g.,}}
\cite{Cugnon1987, Ono2006, Bondorf1976, Stocker1986,Fuchs2006}),
but, summarizing all pros and cons, it is possible to conclude that
although much progress has been made over the past couple of
decades, we are still far from having models that are formally well
founded, practically applicable, and sufficiently realistic to be
quantitatively useful.\,\,The description of nuclear fragmentation
dynamics requires that the proper account for the basic quantal
nature of the system be taken \cite{Ono2006}.

Similar situation is observed for the later stages of the
collision.\,\,One of the considerable problems is that the time
needed for the equilibration and the transition to the statistical
description is still under debate.\,\,The other difficulty to be
overcome is that most statistical multifragmentation codes have been
developed to describe specific sets of data and nearly all of them
have different assumptions.\,\,They are not equivalent
\cite{Tsang2006}.\,\,Basically, two main sets of models are used to
describe the final stages of the reaction that are the
multifragmentation models and the evaporation models.\,\,From the
data available in the literature \cite{Tsang2006}, one can see
(Figs.~\ref{mf2} and \ref{mf4}) that the overall behavior of the
systems shows a similarity among models suggesting some predictive
power as for system's behavior in the multifragmentation phenomena
and allowing for suggesting physical processes responsible for the
formation of fragments.\,\,At the same time, the behaviors of single
observables differ from model to model suggesting some limitations
on their applicability in treating the experimental results.

In spite  of a long history and a high number of different
approaches used, there is still a number of problems left that have
no explanation.\,\,For example, the models involving a phase
transition have the strong position based on the recent works
dealing with the bimodality \cite{Bonnet2009}.\,\,But, in the same
time, their justification is quite questionable.\,\,They are
insensitive to many initial parameters \cite{Hufner1985}, when it is
known for ordinary liquids that the phase transition is usually the
quantity difficult to be prepared.\,\,The most successful models in
describing the observed fragment mass distribution nowadays are the
statistical equilibrium models \cite{Botvina1995, Campi2003} that
try to reduce the intractable problem of a time-dependent highly
correlated interacting many-body fermion system to the much simpler
picture of a system of non-interacting clusters
\cite{Bertsch2004}.\,\,First based on Bevalac's results, they were
used for intermediate energies of HIC with many substantial
variations \cite{Das2003a}.\,\,In the same time, such models have a
number of difficulties \cite{Botvina1995, Hufner1985}.\,\,The
question with the equilibrium at a low freeze-out density that lies
in the basis of such theories is not confirmed by some microscopic
approaches.\,\,There are evidences of only a small part of the
system being heated \cite{Hufner1985, Campi2003}.\,\,The other
problem concerns with the predicted kinetic energy of fragments,
being lower than the temperature of the gas of fragments
\cite{Campi2003},~{\textit{etc.}}

\begin{figure*}
\vskip1mm
\includegraphics[width=14cm]{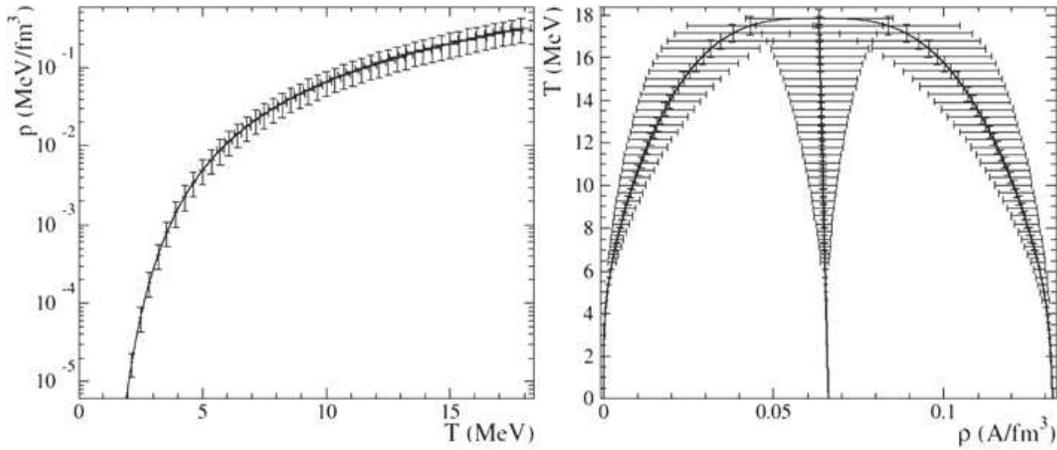}
\vskip-3mm \parbox{14cm}{\caption{Left: the pressure-temperature
coexistence curve for bulk nuclear matter.\,\,Right: the
temperature--density coexistence curve for bulk nuclear matter.
Errors are shown for selected points to give an idea of the error on
the entire coexistence curve \cite{Moretto2011}} \label{coex1}}
\end{figure*}

Therefore, the determination of the mechanism responsible for the
multifragmentation phenomenon and able to solve the existing
problems in natural way is of particular interest as this issue has
not been settled up to now.\vspace*{-2mm}

\subsection{Thermodynamics of heavy ion collisions}
\begin{figure}
\vskip1mm
\includegraphics[width=7.0cm]{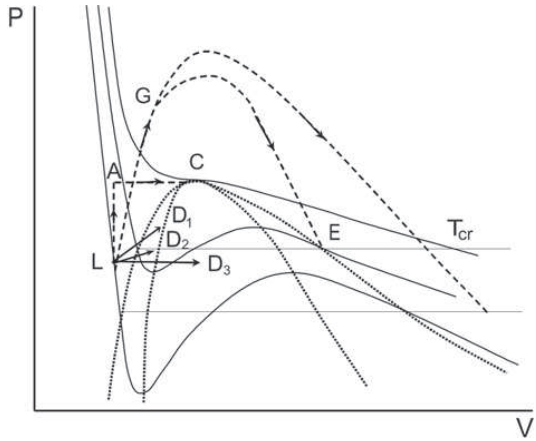}
\vskip-3mm \caption{Possible phase trajectories of a nuclear system
in the proton-induced multifragmentation phenomenon} \label{fig2}
\end{figure}

The interest to the intermediate-energy collisions is triggered by a
fact that the multifragmentation is claimed to be connected with a
phase transition in NM \cite{Moretto2011}.\,\,Such an assumption
comes from the thermodynamic analysis of the coexistence curve in NM
(Fig.\,\,\ref{coex1}) based on the intermediate HIC data.\,\,In that
case, the Fisher droplet model is widely used due to the fact that
the distribution of fragments can be described by Fisher's
well-known formulae for the droplet mass distribution
\cite{Panagiotou1984, Fisher1967}
\begin{equation}
\frac{dP}{dM}\sim{M}^{-\alpha},
\label{eq3}
\end{equation}
where $P$ is the fragment formation probability, $M$ is the mass,
and $\alpha$ is the critical exponent.\,\,In that case, the basic
principles of the analysis are absolutely the same as in the physics
of ordinary liquids.\,\,The description of the formation of clusters
is based \cite{Elliott2013} on the Gibbs
 balance equation for the free energy $G$:
\begin{equation}
\Delta{G}={\Delta}E-T{\Delta}S+P{\Delta}V,
\label{eq4ch1}
\end{equation}
where ${\Delta}E$ and  ${\Delta}S$ are the energy and entropy costs
of the formation of a cluster, respectively, $P$ is the pressure
and ${\Delta}V$ is a change in the volume owing to the formation of
a cluster. Within such approach, it appears next to be possible to
use the classical Guggenheim relation for the coexistence curve
\cite{Guggenheim1945}:
\begin{equation}
\frac{\rho_{l,v}}{\rho_c}=1+b_1\left(\!1-\frac{T}{T_c}\!\right){\pm}b_2\left(\!1-\frac{T}{T_c}\!\right)^{\!\!\beta}\!,
\label{eq5ch1}
\end{equation}
where $b_1$, $b_2$  are the parameters and
$\beta=\frac{(\tau-2)}{\sigma}$ \cite{Fisher1967} is the critical
exponent.\,\,Equation (\ref{eq5ch1}) together with the experimental
vapor branch of the coexistence cur\-ve~\cite{Elliott2013}
\begin{equation}
\frac{\rho}{\rho_c}=\frac{\sum{An_A(T)}}{\sum{An_A(T_c)}},
\label{eq6ch1}
\end{equation}
allow constructing the complete coexistence curve of nuclear matter
({{e.g.}}, see Fig.~\ref{coex1}).\,\,Within this approach, the
modern theory of phase transitions together with the thermodynamics
allow estimating the critical temperature in infinite NM and a
finite nucleus.\,\,That can give, in turn, a deep insight into the
properties of NM and can be of great importance for developing the
proper NM EOS.\,\,Quite promising is the thermodynamic analysis of
the phase trajectories of systems formed in HIC.\,\,Recently, there
appeared a work devoted to the study of different possible phase
trajectories of the excited nuclear system at the $P-V$ plane
\cite{Cherevko2011}.\,\,It was aimed to get a qualitative picture of
the phenomenon considering the boundness of the system and the
existing laws, which describe the behavior of the ordinary liquids
in the metastable and spinodal states
\cite{Scripov1974}.\,\,Assuming the system before the collision to
be at point $L$ of the phase diagram (Fig.~\ref{fig2}), the
different scenarios of system's evolution were studied.

Evolution of the system crucially depends on the ex\-ci\-tation
energy, mass number, and energy release conditions.\,\,This position
suggests two different groups of phase trajectories, namely: the
single-phase transition (marked with the dashed line) and two-phase
transitions (marked with the solid li\-ne).\,\,It is obvious that
there could be a mixture of two decay channels, when different parts
of the system are found in different areas of the phase diagram.
Thus, one may say that the nuclear multifragmentation is a nonlinear
phenomenon, which means that the qualitative picture is different
depending on system's pa\-ra\-me\-ters.\,\,The thermodynamic
analysis conducted in the work has shown that not all of the
mechanisms could be realized in nuclear systems because of their
size.\,\,Some mechanisms does not meet the requirements for the time
needed for the pro\-cess.\,\,Sum\-ming up all pros and cons of
different decay channels and qualitative characteristics, it was
suggested that the spinodal decomposition could not be responsible
for the multifragmentation phenomena because of the system
size.\,\,That fact can be explained from Cahn's theory of separation
by the spinodal decomposition \cite{Cahn1965}.\,\,In that theory,
changes in the free energy read
\begin{equation}
\label{eq9} \Delta F=\int{\left[ \frac{1}{2}\left(\! \frac{\partial
^{2}f}{\partial c^{2}} {\left( c-c_{0} \right)}^{2}
\!\right)+K{\left( \nabla c \right)}^{2} \right]}dV,
\end{equation}
where  $f$ is the free-energy density of a homogeneous material with
composition  $c$, $K{\left( \nabla c \right)}^{2}$  is the
additional free energy density if the material is in a gradient in
the composition.\,\,Therefore, the system is unstable relative to
the Fourier components with $\beta < \beta_{c}={\left(\!
-\frac{{{\partial^{2}}f}/{\partial {{c}^{2}}}\;}{2K}
\!\right)}^{\!\!{1}/{2}}$ or sufficiently large wavelengths, as this
decreases the free energy, when the system is in the unstable
region.\,\,This result shows that, for the smaller wavelengths,
there is no decrease in system's free energy, and there is no sign
change in the diffusion coefficients.\,\,Therefore, for a small
enough system, the spinodal behavior cannot be justified.\,\,As for
the most appropriate decay channel, it appeared to be the mechanical
breakdown of the shell in a single-phase process that may be
followed by the metastable boiling.\,\,Although, there also could be
a mechanism based on the metastable boiling of the inner part of the
system.\,\,The suggested thermodynamic analysis allowed for a simple
qualitative picture of the phenomena.\,\,At the same time, some
further quantitative estimations based on the combination of
macroscopic and microscopic theories, as well as on computer
simulation results, are needed.

There exist quite a lot of other examples (e.g., \cite{Boyko1991}
and references therein) of the successful implementation of
statistical thermodynamics approaches to quantum systems and nuclear
matter, in particular.\,\,This suggests the thermodynamics to be an
important ingredient in the nuclear matter studies.

\subsection{Exotic topologies in the intermediate energy collisions.\,\,Hydrodynamic approaches}
\begin{figure}
\vskip1mm
\includegraphics[width=6.5cm]{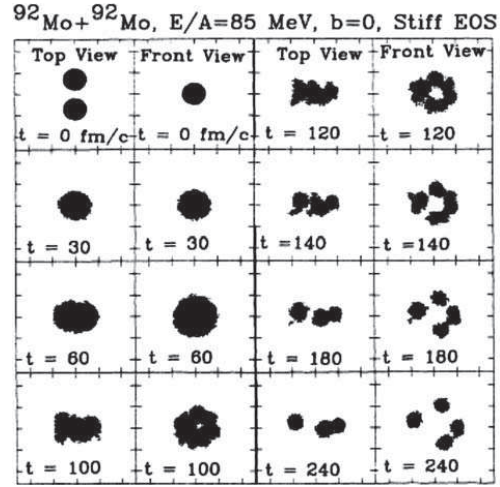}
\vskip-3mm \caption{BUU calculations with the stiff EOS for
$^{92}$Mo~+ $^{92}$Mo collisions at $E/A = 85$~MeV, $b = 0$.\,\,Only
the areas with densities $\rho>0.1\rho_0$ are shown.\,\,The scales
between neighboring ticks are 10~fm \cite{Xu1993}} \label{Topo1}
\end{figure}

\begin{figure}
\vskip1mm
\includegraphics[width=6.5cm]{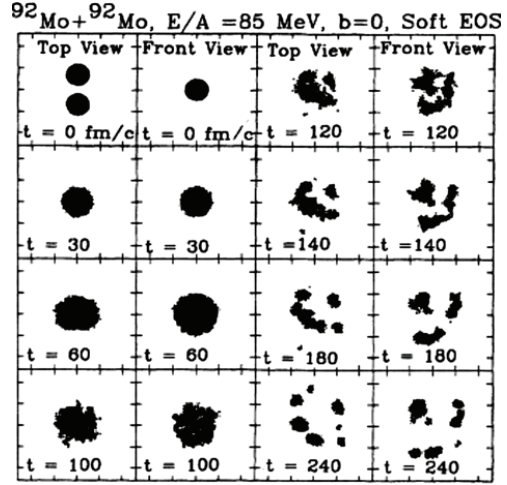}
\vskip-3mm \caption{BUU calculations with the soft EOS for
$^{92}$Mo~+ +~$^{92}$Mo collisions at $E/A = 85$~MeV, $b =
0$.\,\,Only the areas with densities $\rho>0.1\rho_0$ are
shown.\,\,The scales between neighboring ticks are 10~fm
\cite{Xu1993}} \label{Topo2}
\end{figure}

\begin{figure}
\vskip1mm
\includegraphics[width=6.5cm]{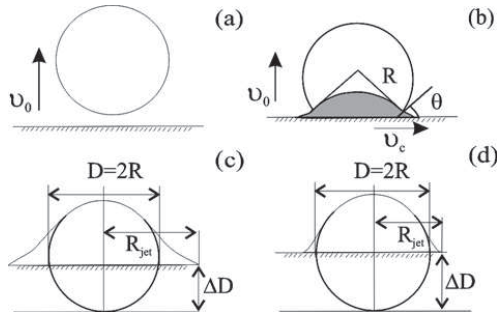}
\vskip-3mm \caption{Different stages of the evolution: symmetry
plane impacting the nuclei \textit{a}; lateral jetting, when the
shockwave velocity becomes equal to the contact edge velocity
\textit{b}.\,\,Compressed zone is shown in grey.\,\,The geometry of
the system in the case of
{\textquotedblleft}stiff{\textquotedblright} EOS ($K=380$~MeV)
\textit{c}; geometry in the case of
{\textquotedblleft}soft{\textquotedblright} EOS ($K=200$~MeV)
\textit{d}} \label{fig1}
\end{figure}

The hydrodynamic description of nuclear matter dates back to the
1980s (see, e.g., \cite{Siemens1979, Stocker1980}) and is now widely
used for the high energy HIC.\,\,In some pioneer works
\cite{Baumgardt1975}, the possibility to use the hydrodynamic
description in the lower-energy limit was confirmed, by basing on
the analysis of the nucleon mean free path that was defined as
$\lambda=1.4\frac{\rho_0}{\rho}$fm.\,\,Later on, quite a lot works
addressed the problem of a nucleon mean free path and the evaluation
of the in-medium nucleon cross section, with the Pauli blocking
being considered \cite{Li1993, Alm1995, Li2006}.\,\,Still, the
results found in some more recent publications \cite{Chen2003} with
the parametrized in-medium nucleon-nucleon cross sections from the
Dirac--Brueckner approach based on the Bonn-A potential give the
values $\lambda\sim1.4$~fm and $\lambda\sim1.3$~fm for
$\rho/\rho_0=1$ and $\rho/\rho_0=1.5,$ respectively, at $E=$
$=50$~MeV/nucleon.\,\,Those results confirm well the argumentation
in \cite{Baumgardt1975}.\,\,Despite that, the hydrodynamic
description of the HIC at intermediate and low energies has been the
subject of only few theoretical studies \cite{Tang1980}.\,\,During
the last decades in a number of works, the possibility of the
formation of toroidal (Fig.\,\,\ref{Topo1}) and bubble
(Fig.\,\,\ref{Topo2}) structures in the head-on HIC was studied
extensively with the help of microscopic transport models
\cite{Bauer1992, Xu1993}.\,\,Within the
Boltzmann--Uehling--Uhlenbeck (BUU) model, it was shown that, at the
energy interval 60--75~MeV/A, the different exotic topologies of the
formed objects can be observed, depending on the stiffness of the
nuclear mater EOS.\,\,Main  peculiarities observed in the
qualitative picture, which comes from the
Boltzmann--Uehling--Uhlenbeck calculations and the available
experiments \cite{Bauer1992, Xu1993}, are the fact that the
expanding velocity of the outer surface of the system is much
smaller than the expanding velocity of the inner surface,
simultaneous breakup with few fragments of similar masses and low
kinetic energies and with the angular distribution of fragments that
is almost isotropic in the case of soft EOS (the incompressibility
at saturation density $K_0=200$~MeV), and the presence of fragments
in the plane perpendicular to the beam direction for the stiff EOS
(the incompressibility at the saturation density
$K_0=380$~MeV).\,\,For some time, the experimental results were not
able to confirm the occurrence of the predicted geometries
\cite{Moretto1997}.\,\,But, later on, the signatures of the
{\textquotedblleft}doughnut\textquotedblright--like structures with
the production of similar-size intermediate-mass fragments were
observed in central HIC \cite{Stone1997} and again confirmed by
transport models \cite{Chen2003}.\,\,Unfortunately the calculations
with transport theories used for the phenomena in focus are not able
to give reliable information on multiparticle observables at the
late stage of the process due to the fact that they do not include
multiparticle correlations and fluctuations \cite{Xu1993}.\,\,In
some works \cite{Bauer1992, Allen1975}, the Rayleigh--Taylor
mechanism was suggested to be responsible for the formation of
fragments, when the exotic topologies were explained with the help
of shock waves.\,\,Such type of structures is also somewhat similar
to those studied theoretically from another approach
\cite{Wong1985}.\,\,Different shapes observed in the experiment may
yield the valuable information about the nuclear incompressibility,
which is a key parameter of the nuclear EOS, as well as the
information regarding the surface tension and NM viscosity.\,\,The
possibility for different shock wave mechanisms to be involved into
the process can also give an extra link in between the
Boltzmann-like transport theory and the hydrodynamic approach.
Recently, it was shown \cite{Cherevko2014} that the observed
qualitative picture has much in common with the collisions of
high-speed ordinary droplets \cite{Bartolo2006, Rioboo2002,
Thoroddsen1998}.\,\,%
For the latter in the certain energy range, the formation of
{\textquotedblleft}doughnut\textquotedblright--like structures with
the following fragmentation into several secondary drops of
approximately equal masses is observed \cite{Pan2009}, which has
much in common with the nuclear case \cite{Xu1993}.\,\,In a recent
work \cite{Cherevko2014}, the system of two identical heavy nuclei
(e.g.\,\,${}^{93}$Nb~+~${}^{93}$Nb) involved in a head-on collision
was considered.\,\,The symmetry of the system allowed the
simplification of the model by changing to the collision of a
spherical nuclei of radius $R$ and density $\rho_0$ with a rigid
wall that moved toward it with the velocity $\upsilon_0$
(Fig.~\ref{fig1},~\textit{a}).\,\,The slip boundary condition was
applied on the wall surface to account for the difference in a
viscous \mbox{behavior.}\looseness=2

\begin{table}[b!]
\vspace*{2mm} \noindent\caption{Quantitative  characteristics\\ of
the process for the \boldmath${}^{93}$Nb + ${}^{93}$Nb system\\ at
60~MeV/nucleon with {\textquotedblleft}stiff{\textquotedblright} EOS
($K=380$~MeV) } \vskip3mm\tabcolsep8.5pt \label{tab2}
\noindent{\footnotesize\begin{tabular}{|l|c|c|c|}
 \hline \multicolumn{1}{|c}
{\rule{0pt}{5mm}} & \multicolumn{1}{|c}{\rule{0pt}{5mm}\cite{Cherevko2014}}& \multicolumn{1}{|c}{\cite{Bauer1992, Xu1993}}& \multicolumn{1}{|c|}{\cite{Stone1997}}\\[2mm]%
\hline%
\rule{0pt}{5mm}Reaction time $\tau_3$, fm/c & 118 & 120--160 & $\;$ \\%
Maximum density ${\rho }/{{{\rho }_{0}}}\;$ & 1.6 & $\sim$1.5& $\;$ \\%
Spreading size $\frac{R_{l}^{\max }}{R}$ & 2.6 & $\sim$2 & $\;$ \\%
Rim radius at $\tau_3$, fm & 1.9 & \; & $\;$ \\%
Number of similar mass IMF & 5.6 & 3--6 & 4--6 \\[2mm]%
\hline
\end{tabular}}
\end{table}

From the analysis of the recent theoretical and experimental results
for the collisions of liquid droplets \cite{Eggers2010, Chizhov2000,
Roisman2012} together with the BUU \cite{Xu1993} and nuclear fluid
dynamics (NFD) \cite{Stocker1986} results for HIC, it is suggested
that there should be four distinct stages of the collision
[Fig.~\ref{fig1}].\,\,Among them are the violent stage at the
beginning of the process, when the highly compressed zone is formed,
and the second stage that is characterized by the lateral jetting
and lasts till the shock wave from collision plane reaches the
boundary of the nuclei.\,\,During those two stages, the final
topology of the system is fully defined.\,\,At the later times, the
system goes through the third stage corresponding to the expansion
process and finally comes to the last stage, when the fragmentation
takes place.\,\,The main quantitative estimations within the
suggested model and their comparison with the BUU calculations and
experimental data are presented in Table.

One can see that the above model slightly overestimates the
spreading size and underestimates the reaction time.\,\,This
probably comes from the approximations used in the model.\,\,For
example, accounting for the viscous dissipation and changing to the
Navier--Stokes equations, as well as accounting for the changes in
the surface tension coefficient due to the temperature increase, can
influence both of the above values.\,\,The model doesn't give the
strict results, but rather provides a picture of the involved
physical processes as clear as possible.\,\,For such a model with no
adjustable parameters, the results for the maximum density and
overall timescale of the process are in good correspondence with the
BUU calculations.\,\,It is also worth mentioning that the result for
the number of fragments of similar masses is in good correspondence
both with the experiment and the BUU calculations.\,\,Such results
show that the hydrodynamic description based on the laws developed
for ordinary liquids seems to explain the overall behavior of the
nuclear systems in the head-on HIC in focus and allows a simple
physical picture of the \mbox{formation} of exotic structures.
\looseness=1

One can conclude that the combination of the hydrodynamic approach
together with the transport theory calculations can reveal the
physical nature of the phenomena observed in HIC and give
possibility to proceed with extracting the data on the NM properties
from the HIC at different energies.

\section{Surface Effects in Nuclear\\ Matter: Macroscopic Description}
The nuclear  matter properties have been extensively studied for
more than half a century.\,\,There exist quite a number of
theoretical works devoted to different models of infinite nuclear
matter, as well as finite nuclei.\,\,When discussing the different
approaches, one can divide them into two groups that are microscopic
and macroscopic approaches.\,\,At first glance, the microscopic
approaches seem to be much more general than the macroscopic ones
that deal only with the averaged values and can be obtained
basically as the average of the microscopic theory.\,\,At the same
time, the exact microscopic calculations are very challenging to
perform, and the precise data on the microscopic parameters are
needed that makes it difficult for practical applications.\,\,During
recent decades, the impressive  progress has been achieved in the
macroscopic description of nuclear matter \cite{Brack1985,
Chomaz2005, Moretto2011}.\,\,A number of papers devoted to the
thermodynamics of small systems or the hydrodynamics of nuclear
matter appeared \cite{ Baumgardt1975, Siemens1979,Wong2008,
Gross2001, Chomaz2004}.\,\,Therefore, the nuclear systems that
consist of a number of nucleons that is neither very small nor very
large seem to be a good object for being studied with the help of a
combination of both macroscopic and microscopic approaches.\,\,The
bright example of such a cooperation of the methods in nuclear
physics is the liquid droplet model \cite{Myers1969} introduced in
the 1960s, which makes possible the description of the average
properties of a saturated system such as a nucleus consisting of two
components (neutrons and protons) with account for the boundary
effects and the presence of a diffuse layer.\,\,In that case, the
energy at equilibrium is given as
\[
E=\left(\!-a_1+J{\bar{\delta}}^2+\frac{1}{2}K{\bar{\epsilon}}{\bar{\delta}}^2+\frac{1}{2}M{\bar{\delta}}^4\!\right)A\,+
\]\vspace*{-7mm}
\[
+\,a_2(1+2{\bar{\epsilon}})A^\frac{2}{3}+Q\tau^2A^\frac{2}{3}+a_3A^\frac{1}{3}\,+
\]\vspace*{-7mm}
\[
+\,c_1\frac{Z^2}{A^\frac{1}{3}}\left(\!1-{\bar{\epsilon}}+\frac{1}{2}{\tau}A^{-\frac{1}{3}}\!\right)-
\]\vspace*{-7mm}
\begin{equation}
\label{eq000ch1}
  -\,c_2Z^2A^\frac{1}{3}-c_3\frac{Z^2}{A}-c_4\frac{Z^\frac{4}{3}}{A^\frac{1}{3}}
\end{equation}
with
\begin{equation}
\label{eq0001ch1}
\begin{array}{l}
 \displaystyle \frac{\rho}{\rho_0}=1-3\epsilon,\\[3mm]
\displaystyle   \frac{\rho_n}{\rho}=\frac{1}{2}(1+\delta)
  \frac{r_0}{R}=A^{-\frac{1}{3}},
  \end{array}
  \end{equation}
where $J$, $K$, $M$, $Q$, $a_1$, $a_2$, and $a_3$ are the coefficients evaluated from the observed averaged phenomenological inputs and some relations based on the different microscopic considerations, $A$ is the nuclear mass number, and $Z$ is the charge number. As a result, the liquid droplet model is able to provide a quite precise description of the properties of finite nuclei. As comes from the liquid droplet model in studying the curvature-correction term for the nuclear matter, one should keep in mind the connection between the surface and bulk properties of the matter \cite{Myers1969, Moretto2012}. As shown in the droplet model, the coefficients in the term proportional to $A^{\frac{1}{3}}$ in the expansion of nuclear properties in terms of the fundamental dimensionless ratio $A^{-\frac{1}{3}}$ are related to the bulk properties of NM described by the terms proportional to $A$ and $A^{\frac{2}{3}}$.

Corrections due to a curvature may play an important role, when
studying light nuclei or processes, where surface terms are
important.\,\,Particularly important are those corrections in the
interpretation of multifragmentation experiments \cite{Gross1990,
Zhang1996, Zhang1998, Chomaz2004}, in which light nuclei necessarily
appear.\,\,The exponential dependence of the yield of fragments on
the surface tension makes this process sensitive to the curvature
corrections \cite{Elliott2013,Toke2003}.\,\,Other important
phenomena that may be affected by changes in the surface tension due
to curvature corrections are:

a) the appearance of the neck region in the fission processes and

b) the hydrodynamic  instability of the structures formed in heavy
ion experiments governed by surface effects \cite{Brosa1983,
Cherevko2014}.

Quite a number of papers devoted to studies of the surface energy
and the  properties of surfaces in NM \cite{Ravenhall1983,
Boyko1990, Jenkovszky1994} have appeared.\,\,Furthermore, the
dependence of the surface tension coefficient (and surface energy)
on the surface curvature and its influence on different physical
properties were also studied by various groups of researchers
\cite{Moretto2012, Pomorski2003}.\,\,Still for decades, it remains a
controversial issue in mesoscopic thermodynamics \cite{Anisimov2007,
Blokhius2006, Kolomietz2012}.

\begin{figure}
\vskip1mm
\includegraphics[width=7.5cm]{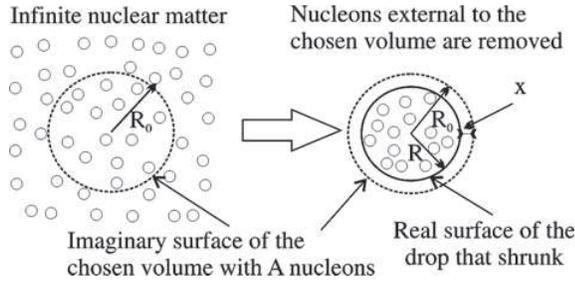}
\vskip-3mm \caption{Schematic picture of the {\it gedanken}
experiment \cite{Cherevko2014a}} \label{fig1a}
\end{figure}

The thermodynamic description of the curvature correction,
originating from the difference between the equimolar surface and
the surface of tension \cite{Rowlinson1982, Rowlinson1994}, dates
back to the 1940s. The Tolman length $\delta$ was originally
introduced in \cite{Tolman1949} to describe the curvature dependence
of the surface tension of a small liquid droplet.\,\,It was defined
as a correction term in the surface tension $\sigma$ of the
liquid-vapour droplet in the isothermal case:
\begin{equation}
\label{eq1}
   \sigma(R)=\sigma_\infty\left(\!1-\frac{2\delta}{R}+...\!\right)\!\!,
 \end{equation}
where $R$ is the droplet radius equal to the radius of the surface
tension \cite{Rowlinson1982, Rowlinson1994}, and $\sigma_\infty$ is
the surface tension of a planar interface.\,\,Eq.\,(\ref{eq1})
originates from the Gibbs--Tolman--Koenig--Buff's thermodynamic
equation and the  assumption that $\delta$ is independent of $R$ for
$\delta\ll{R}$ \cite{Ono1960}.\,\,This physics should work not only
for liquid droplets, but also for any system with curved interface
of a non-negligible boundary layer \cite{Anisimov2007}.\,\,This
situation corresponds to nuclei and nuclear systems with a finite
diffuse layer \cite{Brack1985}.\,\,Based on that fact, a
thermodynamic approach that links the EOS of NM with the curvature
corrections to the surface tension coefficient has been introduced
recently \cite{Cherevko2014a}.\,\,In this work, the analysis is
based on the simple gedanken experiment.\,\,First, one should
consider infinite nuclear matter ($P_0$, $T=$~const) with the chosen
spherical volume $V_0=\frac{4}{3}{\pi}{R_0}^3$ in it consisting of
$A$ nucleons.\,\,If all the nucleons outside the chosen volume are
removed, one gets a {\textquotedblleft}nuclear
droplet{\textquotedblright} that, due to the surface tension,
shrinks to the volume $V=\frac{4}{3}{\pi}{R}^3$, where $R$ is the
final radius of the chosen volume (Fig.~\ref{fig1a}).\,\,This
droplet remains in equilibrium.\,\,The analysis of changes in the
surface size due to the shrinkage allows calculating the curvature
correction to the surface tension coefficient from the EOS of NM.
The Tolman length $\delta$ is calculated for different effective
interactions ({SLy6, SkM*} and {SV-min}).\,\,The calculated results
for $\delta$ in the case of, for example, the $SV$-min
parametrization, appeared to be $\delta\sim-0.55$~fm that is close
to the distance between the nucleons.\,\,That result correspond to
the theoretical estimations based on the Gibbs--Tolman formalism.
Therefore, the statistical thermodynamic approach can be useful in
studies of the surface properties of nuclei.\vspace*{-2mm}

\section{Hadronic Matter and Quantum Liquids}
Nowadays, there exist quite a lot of papers devoted to the
application of classical results of the phase transition theory and
the theory of ordinary liquids for hadronic matter and quantum
liquids. That can be explained by the growing interest in and the
understanding of the universality of critical phenomena in different
media over a tremendous span of substances~-- from ultracold atoms
to dense nuclear matter
\cite{Denicol,Pal,Bozek,Schafer2009}.\,\,Calculations of the
nucleation rate during the (de)confinement phase transition are
subject for continuous debates with implications in the early
Universe and relativistic HIC.\,\,The interest in this subject is
triggered by the interpretation of the observed elliptic flow in
heavy-ion collisions at RHIC \cite{Arsene2005,Back2005,Adcox2005} as
a manifestation of the low shear viscosity of nuclear matter, and a
subsequent realization \cite{Csernai2006} of the universality of
this phenomenon, see also \cite {Gorenstein2008,Khan}.\,\,Most
remarkably, these studies penetrated
\cite{Maeda,Xu,Taylor,Koide,Angilella} different neighboring fields
of physics such as condensed matter and statistical mechanics and
their interface with nuclear and astro-particle
physics.\vspace*{-2mm}

\subsection{Quantum liquids}
When applying the classical  fluid thermodynamics to nuclear matter,
quantum corrections become important as compared to the motion of
atoms and molecules in ordinary liquids \cite{Fisher1964,Boer1948}.
There is an essential difference between nucleons, on the one hand,
and nuclei, atoms, or molecules, on the other hand.\,\,In the latter
case, the relevant energies~-- binding energies, exciting energies,
{\it etc}.\,\,- are much smaller than the rest frame masses of the
system or of the constituents.\,\,In the case of hadrons, they are
of the same order as the mass of the system.\,\,Hence, the quantum
effects are of importance.

The amount  of quantum corrections to the classical theory can be
quantified from the expansion in powers of the Planck constant $h$
of the quantum statistical sum of a system of interacting particles.
For the Helmholtz free energy, the following expression holds
\cite{Fisher1964,Landau1980}, in which the first quantum correction
is proportional to $h^{2}$:
\begin{equation}
\label{eq1a} F = F_{\rm class} \left[ {1 + \left(\! {\frac{{\Lambda}
}{{T^{\ast} }}} \!\right)^{\!\!2}\omega \left( {T^{\ast} ,V^{\ast} }
\right) + ...} \right]\!.
\end{equation}
Here, $F_{\rm class} $ is the classical free energy,  $\omega$ is
some dimensionless function of the order of 1, and a new
dimensionless parameter $\Lambda$ is defined as
\begin{equation}
\label{eq2} \Lambda = \frac{{h}}{{\sigma \sqrt {m\varepsilon} } },
\end{equation}
where $m$ is the mass of the particle. $\Lambda$ is called the de
Boer parameter and describes the amount of quantum effects in
statistical systems \cite{Fisher1964,Landau1980}.

The apparent similarity of the properties of different media
suggests the use of the law of corresponding states (LCS)
\cite{Fisher1964}.\,\,For real atoms and molecules, the
intermolecular potential $\Phi(r)$ can be reduced, with reasonable
precision, to a product of the energy constant $\varepsilon $ (the
depth of the minimum in the particle interaction potential) and a
function of the dimensionless variable $x = \frac{r}{\sigma}$,
\begin{equation}
\label{eq01} \Phi(r)=\varepsilon\phi\left(x\right)\!,
\end{equation}
where $\sigma $  is the position of the minimum.\,\,Then, by LCS,
for a group of substances with a common potential $\phi(x)$, the
universal functions for the pressure $p^{\ast}=p^{\ast} \left(
T^{\ast},V^{\ast} \right) $, energy $E^{\ast}=E^{\ast} \left(
T^{\ast},V^{\ast} \right)$, free energy $F^{\ast}=F^{\ast} \left(
T^{\ast},V^{\ast} \right)$, \textit{etc.} should exist, where
$T^{\ast}=\frac{kT}{\varepsilon}$ is the reduced temperature and
$V^{\ast}=\frac{V}{\sigma^{3}}$ is the reduced volume.\,\,As a
consequence, the thermodynamic properties of all the substances from
a given group are similar in terms of a properly chosen scale.\,\,In
view of the similarity of the nuclear forces to the van der Waals
forces, see Fig.~\ref{fig1}, LCS should work both in ordinary
liquids and in nuclear matter.

\begin{figure}
\vskip1mm
\includegraphics[width=6.0cm]{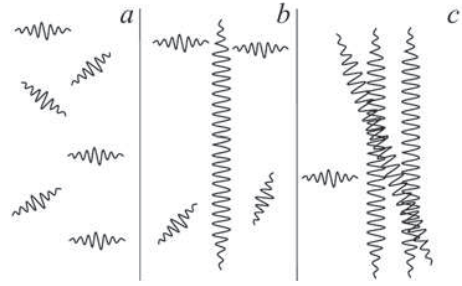}
\vskip-3mm \caption{Distribution of waves of particles: in a
classical system (high temperatures) (\textit{a}); exactly at the
transition temperature (\textit{b}); slightly below the transition
point (\textit{c})} \label{fig3}
\end{figure}

Recently, it was suggested to extend the principle of corresponding
states for ordinary liquids to the possible fluid-superfluid and
conductor-superconductor transitions for a variety of systems,
ranging from  $^3$He,\ \ $^4$He, and superconducting electron fluids
to nuclear matter.\,\,Such an extension is based on the description
of the superfluidity phenomenon in terms of wave packets.
Superfluidity assumes the appearance of phased waves (wave packets)
of bosonic quantum particles due to the Bose condensation (Fig.
\ref{fig3}).

As a consequence, it appeared to be possible to define the analogue
of the de Boer parameter~-- dimensionless number BJEMS
\cite{Bulavin2008}
\begin{equation}
\label{eq8} {\rm BJEMS} = g\, l\left( {mkT} \right)^{1/2}/\hbar.
\end{equation}
Analogously to the de Boer parameter, it takes the quantum effects
into account.\,\,The difference is that it considers the exchange
effects.\,\,Therefore, it might be used for the superfluid phase
transitions.\,\,The parameter BJEMS is assumed to be of the same
order of magnitude for all fluid-superfluid and
conductor-superconductor transitions.\,\,This hypothesis is based on
the general principles, rather than on the microscopic theory, and
it appears to be correct for all the real Fermi and Bose systems
from the same universality class.

\subsection{Hadronic matter}
Studies of the hadronic matter  are an important part of the modern
astrophysics and HIC physics.\,\,In the early Universe, a particular
nucleation dynamics during the confinement can influence the
inhomogeneities in the hadron distribution.\,\,Nowadays, the
possibility for a number of colored objects such as quarks and
gluons to escape the hadronization during the cooling of the
Universe is intensively studied
\cite{Witten1984,Chandra2000,Brilenkov2014}.\,\,This requires the
understanding of the surface properties of the formed quark
{\textquotedblleft}nuggets{\textquotedblright}.\,\,Statistical
thermodynamics can be quite helpful in the studies of those
properties.\,\,In HIC, superheated hadron matter can produce extra
entropy and might lead to observable consequences, detectable in the
experiment.\,\,Earlier, there were the attempts to use statistical
thermodynamic approaches for studying the hadron matter,
confinement, phase transitions, and early evolution of the Universe
\cite{Boyko1990,Jenkovszky1990,Boyko1990a,Bulavin2010a}.\,\,Within
the suggested approach, some interesting results were obtained in
the description of the expansion of the Universe during the early
stages of its evolution.\,\,The behavior of the surface tension
coefficient in the (de)confinement phase transitions was studied.
The obtained results for a parton gas can be useful in the studies
of the color-glass condensate that is discussed in the framework of
the studies of relativistic HIC.

The similarity of the behavior  of ordinary liquids and hadronic
matter, as well as the existence of a number of interesting results
obtained for hadronic matter from the statistical thermodynamic
approaches, suggests the application of classical laws in high-energy physics.

\section{Summary}

Macroscopic theories well-developed for ordinary liquids have shown
themselves to be successful in describing the average properties of
nuclear systems.\,\,They work well for nuclear matter in combination
with microscopic approaches naturally complementing each other.
Hydrodynamic approaches developed for ordinary liquids are widely
used in the studies of heavy ion collisions at different energies,
as they explain well the observed phenomena, and their results can
be easily compared to the calculations within transport theories.
The thermodynamic description has shown to be fruitful for finite
nuclei and intermediate-energy heavy ion collisions.

At the same time, statistical  thermodynamics can be applied not
only for ordinary liquids and nuclear matter, but also for studies
of the properties of quark-gluon plasma.\,\,It has shown to be
useful in understanding the behavior of quantum liquids in phase
transitions.

Macroscopic theories can  be used in constructing a proper equation
of state of nuclear matter that is one of the long standing and
desirable goals in nuclear physics.\,\,They can be used in the
studies of the intermediate heavy-ion collisions, which are a window
into the nuclear matter equation of state and can be treated as the
analogue of the $P$-$V$-$T$ studies in the physics of ordinary
liquids.

Statistical  thermodynamics and phase transition theory widely used
in the physics of ordinary liquids work well for hadronic
matter.\,\,They are quite fruitful in studying the (de)confinement
phase transitions.\,\,Interesting results are found in the works
devoted to the color glass condensate that is discussed in the
context of relativistic heavy ion collisions.\,\,They can be useful
in studies of the early evolution of the Universe.

As a summary, one can conclude that the approaches originally developed
for ordinary liquids are widely used in the studies of hadronic
matter and can give valuable results.

\vspace*{-5mm}
\rezume{%
К.В.\,Черевко, Л.Л.\,Єнковський,\\ В.М.\,Сисоєв, Фен-Шоу Жанг}
{ЗАГАЛЬНИЙ ПІДХІД\\ ДЛЯ ОПИСУ КЛАСИЧНИХ РІДИННИХ\\ СИСТЕМ І ЯДЕРНОЇ
МАТЕРІЇ } {В роботі робиться спроба короткого огляду можливостей
застосування статистичної термодинаміки при вивченні адронної
матерії. Розглядається можливість застосування гідродинаміки для
опису формування різних топологічних структур в лобових зіткненнях
важких іонів за малих енергій. Показано, що такий підхід дозволяє
отримати аналітичний опис кожної стадії процесу та визначити
властивості ядерної матерії (поверхневий натяг, стистливість та ін.)
з аналізу властивостей утворених фрагментів. Розглядаються переваги
термодинамічного аналізу фазових траєкторій систем, що утворюються в
зіткненнях важких іонів. Розглядається можливість використання
статистичної термодинаміки для визначення поправок до коефіцієнту
поверхневого натягу, викликаних кривизною поверхні, з рівняння стану
ядерної матерії.  Розглянуто приклади застосування статистичної
термодинаміки в дослідженнях адронної матерії та квантових рідин.}


\begin{thebibliography}{100}

\bibitem{Bulavin2010}
L.~A. Bulavin and V.~M. Sysoev.
\newblock {\em Physics of phase transitions}.
\newblock VPC "University of Kyiv", Kyiv, 2010.

\bibitem{Bulavin2006}
L.~A. Bulavin, D.~A. Gavryushenko, and V.~M. Sysoev.
\newblock {\em Molecular Physics}.
\newblock Znannya, Kyiv, 2006.

\bibitem{Bulavin2005}
L.~A. Bulavin and V.~K. Tartakovskiy.
\newblock {\em Nuclear Physics}.
\newblock Znannya, Kyiv, 2005.

\bibitem{Gross2001}
D.~H.~E. Gross.
\newblock {\em Microcanonical Thermodynamics: Phase Transitions in 'Small'
  Systems}, volume~66 of {\em World Scientific Lecture Notes in Physics}.
\newblock World Scientific Pub Co Inc, 2001.

\bibitem{Bertsch1988}
G.F. Bertsch and S.~Das Gupta.
\newblock A guide to microscopic models for intermediate energy heavy ion
  collisions.
\newblock {\em Phys. Rep.}, 160(4):189, March 1988.

\bibitem{Baldo2007}
M.~Baldo and C.~Maieron.
\newblock Equation of state of nuclear matter at high baryon density.
\newblock {\em J. Phys. G: Nucl. Part. Phys.}, 34:R243, 2007.

\bibitem{Abe1996}
Y.~Abe, S.~Ayik, P.-G. Reinhard, and E.~Suraud.
\newblock On stochastic approaches of nuclear dynamics.
\newblock {\em Phys. Rep.}, 275:49 -- 196, 1996.

\bibitem{Baran2005}
V.~Baran, M.~Colonna, V.~Greco, and M.~Di Toro.
\newblock Reaction dynamics with exotic nuclei.
\newblock {\em Phys. Rep.}, 410(5):335 -- 466, 2005.

\bibitem{Li2008a}
Bao-An Li, Lie-Wen Chen, and Che~Ming Ko.
\newblock Recent progress and new challenges in isospin physics with heavy-ion
  reactions.
\newblock {\em Phys. Rep.}, 464(4):113 -- 281, 2008.

\bibitem{Karnaukhov2006}
V.A. Karnaukhov.
\newblock Nuclear multifragmentation and phase transitions in hot nuclei.
\newblock {\em Phys. Elem. Part. Atom. Nucl.}, 37:312, 2006.

\bibitem{Shlomo2012}
S.~Shlomo.
\newblock Equation of state of symmetric and asymmetric nuclear matter at
  various densities and temperatures.
\newblock {\em J. Phys: Conf. Ser.}, 337:012014, 2012.

\bibitem{Stevenson2012}
P.D. Stevenson and P.M. Goddard.
\newblock Constraints on skyrme force parameterizations.
\newblock In A.~Georgieva and N.~Minkov, editors, {\em Proceedings of the 31st
  International Workshop on Nuclear Theory}, volume~31. Heron Press, Sofia,
  2012.

\bibitem{Oyamatsu2000}
K.~Oyamatsu, I.~Tanihata, Y.~Sugahara, H.~Toki, and K.~Sumiyoshi.
\newblock Focused on models and theories of the nuclear mass can we determine
  the equation of state of asymmetric nuclear matter using unstable nuclei?
\newblock {\em RIKEN Review}, 26, 2000.

\bibitem{Tsang2012}
M.B. Tsang, J.R. Stone, F.~Camera, P.~Danielewicz, S.~Gandolfi, K.~Hebeler,
  C.J. Horowitz, Jenny Lee, W.G. Lynch, Z.~Kohley, R.~Lemmon, P.~Moller,
  T.~Murakami, S.~Riordan, X.~Roca-Maza, F.~Sammarruca, A.W. Steiner,
  I.~Vidana, and S.J. Yennello.
\newblock Constraints on the symmetry energy and neutron skins from experiments
  and theory.
\newblock {\em arXiv:12040466 [nucl-ex]}, 2012.

\bibitem{Chen2005}
L.W. Chen, C.M. Ko, and B.A. Li.
\newblock Determination of the stiffness of the nuclear symmetry energy from
  isospin diffusion.
\newblock {\em Phys Rev Lett.}, 94:032701, 2005.

\bibitem{Chen2007}
L.W. Chen, C.M. Ko, and B.A. Li.
\newblock Isospin-dependent properties of asymmetric nuclear matter in
  relativistic mean field models.
\newblock {\em Phys. Rev. C}, 76:054316, 2007.

\bibitem{Klahn2006}
T.~Klahn, D.~Blaschke, S.~Typel, E.~N.~E. van Dalen, A.~Faessler, C.~Fuchs,
  T.~Gaitanos, H.~Grigorian, A.~Ho, E.~E. Kolomeitsev, M.~C. Miller, G.~Ropke,
  J.~Trumper, D.~N. Voskresensky, F.~Weber, and H.~H. Wolter.
\newblock Constraints on the high-density nuclear equation of state from the
  phenomenology of compact stars and heavy-ion collisions.
\newblock {\em Phys.Rev. C}, 74:035802, 2006.

\bibitem{Botvina2009}
A.S. Botvina and I.N. Mishustin.
\newblock {\em Phys. At. Nucl.}, 71:1088, 2009.

\bibitem{Souza2009}
S.R. Souza, B.V. Carlson, R.~Donangelo, W.G. Lynch, A.W. Steiner, and M.B.
  Tsang.
\newblock Statistical multifragmentation model with skyrme effective
  interactions.
\newblock {\em Phys. Rev. C}, 79:054602, 2009.

\bibitem{Danielewicz2002}
P.~Danielewicz, R.A. Lacey, and W.G. Lynch.
\newblock {\em Science}, 298:1592, 2002.

\bibitem{Hufner1985}
J.~Hufner.
\newblock Heavy fragments produced in proton-nucleus and nucleus-nucleus
  collisions at relativistic energies.
\newblock {\em Phys. Rep.}, 125:129, 1985.

\bibitem{Botvina2006}
A.S. Botvina and I.N. Mishustin.
\newblock {\em Eur. Phys. J.}, 30:121, 2006.

\bibitem{Kwiatkowski1994}
K.~Kwiatkowski, W.A. Friedman, L.W. Woo, V.E. Viola, E.C. Pollacco, C.Volant,
  and S.J. Yennello.
\newblock Energy dissipation and multifragment decay in the he3+natag system.
\newblock {\em Phys. Rev. C}, 49:1516, 1994.

\bibitem{Pan1995}
J.~Pan and S.~Das Gupta.
\newblock {\em Phys. Rev. C}, 51:1384, 1995.

\bibitem{Viola2004}
V.E. Viola, K.~Kwiatkowski, J.B. Natowitz, and S.J. Yennello.
\newblock Breakup densities of hot nuclei.
\newblock {\em Phys. Rev. Lett.}, 93:132701, 2004.

\bibitem{Chase1995}
K.C. Chase and A.Z. Mekjian.
\newblock {\em Phys. Rev. Lett.}, 75:4732, 1995.

\bibitem{Fuchs2006}
C.~Fuchs and H.H. Wolter.
\newblock Modelization of the eos.
\newblock {\em Eur. Phys. J. A}, 30:5, 2006.

\bibitem{Das2003a}
C.B. Das, S.~Das Gupta, and A.Z. Mekjian.
\newblock {\em Phys. Rev. C}, 67:064607, 2003.

\bibitem{Chaudhuri2009}
G.~Chaudhuri and S.~Das Gupta.
\newblock {\em Phys. Rev. C}, 80:044609, 2009.

\bibitem{Siemens1983}
P.J. Siemens.
\newblock {\em Nature}, 305:410, 1983.

\bibitem{Rodrigues2011}
M.~R.~D. Rodrigues, R.~Wada, K.~Hagel, M.~Huang, Z.~Chen, S.~Kowalski,
  T.~Keutgen, J.~B. Natowitz, M.~Barbui, A.~Bonasera, K.~Schmidt, J.~Wang,
  L.~Qin, T.~Materna, and P.~K. Sahu.
\newblock Neutron multiplicity from primary hot fragments produced in heavy ion
  reactions near fermi energy.
\newblock {\em J. Phys.: Conf. Ser.}, 312:082009, 2011.

\bibitem{Borderie2008}
B.~Borderie and M.F. Rivet.
\newblock {\em Prog. Part. Nucl. Phys.}, 6:551, 2008.

\bibitem{Marie1998}
N.~Marie and et. al.
\newblock Experimental determination of fragment excitation energies in
  multifragmentation events.
\newblock {\em Phys. Rev. C}, 58:256, 1998.

\bibitem{Jaqaman1991}
H.R. Jaqaman and D.H.E. Gross.
\newblock {\em Nucl. Phys. A}, 524:321, 1991.

\bibitem{Campi1986}
X.~Campi.
\newblock {\em J. Phys. A}, 524:321, 1986.

\bibitem{Berkenbusch2001}
M.~Kleine Berkenbusch, W.~Bauer, K.~Dillman, S.~Pratt, L.~Beaulieu,
  K.~Kwiatkowski, T.~Lefort, W.~c.~Hsi, V.~E. Viola, S.~J. Yennello, R.~G.
  Korteling, and H.~Breuer.
\newblock Event-by-event analysis of proton-induced nuclear multifragmentation:
  Determination of the phase transition universality class in a system with
  extreme finite-size constraints.
\newblock {\em Phys. Rev. Lett.}, 88:022701, 2001.

\bibitem{Ono2006}
A.~Ono and J.~Randrup.
\newblock Dynamical models for fragment formation.
\newblock {\em Eur. Phys. J. A}, 30:109, 2006.

\bibitem{Cugnon1987}
J.~Cugnon.
\newblock {\em Nucl. Phys. A}, 462:751, 1987.

\bibitem{Bondorf1976}
J.P. Bondorf, H.T. Feldmeier, and S.~Garpman~E.C. Halbert.
\newblock {\em Phys. Lett. B}, 65:217, 1976.

\bibitem{Stocker1986}
Horst Stocker and Walter Greiner.
\newblock High energy heavy ion collisions - probing the equation of state of
  highly excited hardronic matter.
\newblock {\em Phys. Rep.}, 137(5):277 -- 392, 1986.

\bibitem{Tsang2006}
M.B. Tsang, R.~Bougault, R.~Charity, D.~Durand, W.A. Friedman, F.~Gulminelli,
  A.~Le Fevre, Al.H.Raduta, Ad.R. Raduta, S.~Souza, W.~Trautmann, and R.~Wada.
\newblock Comparisons of statistical multifragmentation and evaporation models
  for heavy-ion collisions.
\newblock {\em Eur. Phys. J. A}, 30(1):129--139, 2006.

\bibitem{Bonnet2009}
E.~Bonnet and ALADIN~Collaborations) et.~al. (INDRA.
\newblock {\em Phys. Rev. Lett.}, 103:072701, 2009.

\bibitem{Botvina1995}
A.S. Botvina, I.N. Mishustin, M.~Begemann-Blaich, J.~Hubele, G.~Imme, I.~Iori,
  P.~Kreutz, G.J. Kunde, W.D. Kunze, V.~Lindenstruth, U.~Lynen, A.~Moroni,
  W.F..J Muller, C.A. Ogilvie, J.~Pochodzalla, G.~Raciti, Th. Rubehn, H.~Sann,
  A.~Schttauf, W.~Seidel, W.~Trautmann, and A.~Werner.
\newblock Multifragmentation of spectators in relativistic heavy-ion reactions.
\newblock {\em Nucl. Phys. A}, 584:737, 1995.

\bibitem{Campi2003}
X.~Campi, H.~Krivine, E.~Plagnol, and N.~Sator.
\newblock {\em Phys.Rev. C}, 67:044610, 2003.

\bibitem{Bertsch2004}
G.F. Bertsch.
\newblock {\em Am. J. Phys.}, 72:983, 2004.

\bibitem{Moretto2011}
L~G Moretto, J~B Elliott, L~Phair, and P~T Lake.
\newblock The experimental liquidвЂ“vapor phase diagram of bulk nuclear matter.
\newblock {\em Journal of Physics G: Nuclear and Particle Physics},
  38(11):113101, 2011.

\bibitem{Panagiotou1984}
A.~D. Panagiotou, M.~W. Curtin, H.~Toki, D.~K. Scott, and P.~J. Siemens.
\newblock Experimental evidence for a liquid-gas phase transition in nuclear
  systems.
\newblock {\em Phys. Rev. Lett.}, 52:496, 1984.

\bibitem{Fisher1967}
M.E. Fisher.
\newblock {\em Physics}, 3:255, 1967.

\bibitem{Elliott2013}
J.~B. Elliott, P.T. Lake, L.G. Moretto, and L.~Phair.
\newblock Determination of the coexistence curve, critical temperature,
  density, and pressure of bulk nuclear matter from fragment emission data.
\newblock {\em Phys. Rev. C}, 87:054622, 2013.

\bibitem{Guggenheim1945}
E.A. Guggenheim.
\newblock The principle of corresponding states.
\newblock {\em J. Chem. Phys.}, 13:253, 1945.

\bibitem{Cherevko2011}
K.~V. Cherevko, L.~A. Bulavin, and V.~M. Sysoev.
\newblock Thermodynamic analysis of multifragmentation phenomena.
\newblock {\em Phys. Rev. C}, 84(4):044603, 2011.

\bibitem{Scripov1974}
V.~P. Scripov.
\newblock {\em Metastable Liquids}.
\newblock Wiley, New York, 1974.

\bibitem{Cahn1965}
J.W. Cahn.
\newblock {\em J. Chem. Phys.}, 42:93, 1965.

\bibitem{Boyko1991}
V.~G. Boyko, L.~L. Jenkovszky, and V.~M. Sysoev.
\newblock {\em EChAYa}, 22:675, 1991.

\bibitem{Siemens1979}
Philip~J. Siemens and John~O. Rasmussen.
\newblock Evidence for a blast wave from compressed nuclear matter.
\newblock {\em Phys. Rev. Lett.}, 42(14):880, April 1979.

\bibitem{Stocker1980}
H.~Stocker, R.Y. Cusson, J.A. Maruhn, and W.~Greiner.
\newblock Medium energy collisions of heavy nuclei in the three-dimensional
  nuclear fluid dynamical-(nfd) and time-dependent hartree-fock (tdhf) models.
\newblock {\em Z. Phys. A}, 294:125--135, 1980.

\bibitem{Baumgardt1975}
H.G. Baumgardt, J.~U. Schott, Y.~Sakamoto, E.~Schopper, H.~Stocker, J.~Hofmann,
  W.~Scheid, and W.~Greiner.
\newblock Shock waves and mach cones in fast nucleus-nucleus collisions.
\newblock {\em Z. Phys. A}, 273:359--371, 1975.

\bibitem{Li1993}
G.Q. Li and R.~Machleidt.
\newblock Microscopic calculation of in-medium nucleon-nucleon cross sections.
\newblock {\em Phys. Rev. C}, 48:1702, 1993.

\bibitem{Alm1995}
T.~Alm, G.~Ropke, W.~Bauer, F.~Daffin, and M.~Schmidt.
\newblock The in-medium nucleon-nucleon cross section and buu simulations of
  heavy-ion reactions.
\newblock {\em Nucl. Phys. A}, 587:815, 1995.

\bibitem{Li2006}
Q.~Li, Z.~Li, S.~Soff, M.~Bleicher, and H.~Stocker.
\newblock Medium modifications of the nucleon-nucleon elastic cross section in
  neutron-rich intermediate energy hics.
\newblock {\em J. Phys. G: Nucl. Part. Phys.}, 32:407, 2006.

\bibitem{Chen2003}
L.~W. Chen, V.~Greco, C.~M. Ko, and B.~A. Li.
\newblock {\em Phys.Rev. C}, 68:014605, 2003.

\bibitem{Tang1980}
Henry H.~K. Tang and Cheuk-Yin Wong.
\newblock Exactly central heavy-ion collisions by nuclear hydrodynamics.
\newblock {\em Phys. Rev. C}, 21:1846, 1980.

\bibitem{Bauer1992}
W.~Bauer, G.~F. Bertsch, and H.~Schulz.
\newblock Bubble and ring formation in nuclear fragmentation.
\newblock {\em Phys. Rev. Lett.}, 69(13):1888, September 1992.

\bibitem{Xu1993}
H.~M. Xu, J.~B. Natowitz, C.~A. Gagliardi, R.~E. Tribble, C.~Y. Wong, and W.~G.
  Lynch.
\newblock Formation and decay of toroidal and bubble nuclei and the nuclear
  equation of state.
\newblock {\em Phys. Rev. C}, 48(2):933, August 1993.

\bibitem{Moretto1997}
L.~G. Moretto, K.~Tso, and G.~J. Wozniak.
\newblock {\em Phys. Rev. Lett.}, 78:824, 1997.

\bibitem{Stone1997}
N.~T.~B. Stone, O.~Bjarki, E.~E. Gualtieri, S.~A. Hannuschke, R.~Lacey,
  J.~Lauret, W.~J. Llope, D.~J. Magestro, R.~Pak, A.~M.~Vander Molen, G.~D.
  Westfall, and J.~Yee.
\newblock Evidence for the decay of nuclear matter toroidal geometries in
  nucleus-nucleus collisions.
\newblock {\em Phys. Rev. Lett.}, 78:2084, 1997.

\bibitem{Allen1975}
R.F. Allen.
\newblock The role of surface tension in splashing.
\newblock {\em J. Colloid Interface Sci.}, 51(2):350, 1975.

\bibitem{Wong1985}
Cheuk-Yin Wong.
\newblock Hot toroidal and bubble nuclei.
\newblock {\em Phys. Rev. Lett.}, 55(19):1973, November 1985.

\bibitem{Cherevko2014}
Konstantin Cherevko, Jun Su, Leonid Bulavin, Vladimir Sysoev, and Feng-Shou
  Zhang.
\newblock "doughnut" nuclear shapes in head-on heavy ion collisions.
\newblock {\em Phys. Rev. C}, 89:014618, 2014.

\bibitem{Bartolo2006}
Denis Bartolo, Christophe Josserand, and Daniel Bonn.
\newblock Singular jets and bubbles in drop impact.
\newblock {\em Phys. Rev. Lett.}, 96:124501, 2006.

\bibitem{Rioboo2002}
R.~Rioboo, M.~Marengo, and C.~Tropea.
\newblock Time evolution of liquid drop impact onto solid, dry surfaces.
\newblock {\em Exp. Fluid.}, 33:112--124, 2002.

\bibitem{Thoroddsen1998}
S.~T. Thoroddsen and J.~Sakakibara.
\newblock Evolution of the fingering pattern of an impacting drop.
\newblock {\em Phys. Fluid.}, 10:1359, 1998.

\bibitem{Pan2009}
Kuo-Long Pan, Ping-Chung Chou, and Yu-Jen Tseng.
\newblock Binary droplet collision at high weber number.
\newblock {\em Phys. Rev. E}, 80:036301, 2009.

\bibitem{Eggers2010}
Jens Eggers, Marco~A. Fontelos, Christophe Josserand, and Stephane Zaleski.
\newblock Drop dynamics after impact on a solid wall: Theory and simulations.
\newblock {\em Phys. Fluid.}, 22:062101, 2010.

\bibitem{Chizhov2000}
A.~V. Chizhov and A.~A. Shmidt.
\newblock Impact of a high-velocity drop on an obstacle.
\newblock {\em J. Tech. Phys.}, 45:18, 2000.

\bibitem{Roisman2012}
Ilia~V. Roisman, Carole Planchette, Elise Lorenceau, and GГјnter Brenn.
\newblock Binary collisions of drops of immiscible liquids.
\newblock {\em J. Fluid Mech.}, 690:512--535, 2012.

\bibitem{Brack1985}
M.~Brack, C.~Guet, and H.-B. Hakansson.
\newblock Selfconsistent semiclassical description of average nuclear
  properties - a link between microscopic and macroscopic models.
\newblock {\em Phys. Rep.}, 123:275--364, 1985.

\bibitem{Chomaz2005}
Ph. Chomaz, F.~Gulminelli, and O.~Juillet.
\newblock Generalized gibbs ensembles for time dependent processes.
\newblock {\em Annals Phys.}, 320:135--163, 2005.
\newblock Интересно посмотреть.

\bibitem{Wong2008}
C.-Y. Wong.
\newblock Landau hydrodynamics reexamined.
\newblock {\em Phys. Rev. C}, 78:054902, 2008.

\bibitem{Chomaz2004}
Philippe Chomaz, Maria Colonna, and Jorgen Randrup.
\newblock Nuclear spinodal fragmentation.
\newblock {\em Phys. Rep.}, 389:263, 2004.

\bibitem{Myers1969}
William~D. Myers and W.J. Swiatecki.
\newblock Average nuclear properties.
\newblock {\em Ann. Phys.}, 55:395--505, 1969.

\bibitem{Moretto2012}
L.~G. Moretto, J.~B. Elliott, P.~T. Lake, and L.~Phair.
\newblock New wrinkles on an old model: Correlation between liquid drop
  parameters and curvature term.
\newblock {\em AIP Conf. Proc.}, 1491:75, 2012.

\bibitem{Gross1990}
D.~H.~E. Gross.
\newblock Statistical decay of very hot nuclei-the production of large
  clusters.
\newblock {\em Rep. Prog. Phys}, 53:605, 1990.

\bibitem{Zhang1996}
Feng-Shou Zhang.
\newblock Phase transitions, correlations and fluctuations of nuclear
  multifragmentation.
\newblock {\em Zeitschrift fГјr Physik A Hadrons and Nuclei}, 356(1):163--170,
  1996.

\bibitem{Zhang1998}
F.-S. Zhang and L.~X. Ge.
\newblock {\em Nuclear Multifragmentation}.
\newblock Beijing: Science Press, 1998.

\bibitem{Toke2003}
Jan Toke, Jun Lu, and W.~Udo Schroder.
\newblock Surface entropy in statistical emission of massive fragments from
  equilibrated nuclear systems.
\newblock {\em Phys.Rev. C.}, 67:034609, 2003.

\bibitem{Brosa1983}
U.~Brosa and S.~Grossmann.
\newblock In the exit channel of nuclear fission.
\newblock {\em Z. Phys. A}, 310:177, 1983.

\bibitem{Ravenhall1983}
D.G. Ravenhall, C.J. Pethick, and J.M. Lattimer.
\newblock Nuclear interface energy at finite temperatures.
\newblock {\em Nucl. Phys. A}, 407:571, 1983.

\bibitem{Boyko1990}
V.~G. Boyko, L.~L. Jenkovszky, and V.~M. Sysoev.
\newblock Phase transitions in nuclear matter: metastability and fluctuations.
\newblock {\em Zeitsch. Phys. C}, 45:607, 1990.

\bibitem{Jenkovszky1994}
L.~L. Jenkovszky, B.~Kampfer, and V.~M. Sysoev.
\newblock Bubble free energy during the confinement transition.
\newblock {\em Sov. J. Nucl. Phys.}, 57:1507, 1994.

\bibitem{Pomorski2003}
K.~Pomorski and J.~Dudek.
\newblock Nuclear liquid-drop model and surface-curvature effects.
\newblock {\em Phys. Rev. C}, 67:044316, 2003.

\bibitem{Anisimov2007}
M.~A. Anisimov.
\newblock Divergence of tolman’s length for a droplet near the critical point.
\newblock {\em Phys. Rev. Lett.}, 98:035702, 2007.

\bibitem{Blokhius2006}
E.~M. Blokhius and J.~Kuipers.
\newblock Thermodynamic expressions for the tolman length.
\newblock {\em J. Chem. Phys.}, 124:074701, 2006.

\bibitem{Kolomietz2012}
V.~M. Kolomietz, S.~V. Lukyanov, and A.~I. Sanzhur.
\newblock Curved and diffuse interface effects on the nuclear surface tension.
\newblock {\em Phys. Rev. C}, 86:024304, 2012.

\bibitem{Rowlinson1982}
J.~S. Rowlinson and B.~Widom.
\newblock {\em Molecular Theory of Capillarity}.
\newblock Clarendon, Oxford, 1982.

\bibitem{Rowlinson1994}
J.~S. Rowlinson.
\newblock A drop of liquid.
\newblock {\em J. Phys.: Condens. Matter}, 6:A1, 1994.

\bibitem{Tolman1949}
Richard~C. Tolman.
\newblock The effect of droplet size on surface tension.
\newblock {\em J. Chem. Phys.}, 17(3):333, March 1949.

\bibitem{Ono1960}
S.~Ono and S.~Kondo.
\newblock {\em Molecular Theory of Surface Tension in Liquids}.
\newblock Springer-Verlag, 1960.

\bibitem{Cherevko2014a}
K.~V. Cherevko, L.~A. Bulavin, L.~L. Jenkovszky, V.~M. Sysoev, and Feng-Shou
  Zhang.
\newblock Evaluation of the curvature-correction term from the equation of
  state of nuclear matter.
\newblock {\em Phys. Rev. C}, 90:017303, 2014.

\bibitem{Denicol}
G.S. Denicol, T.~Kodama, and T.~Koide.
\newblock {\em arXiv:1002.2394 [nucl-th]}.

\bibitem{Pal}
Subrata Pal.
\newblock {\em arXiv:1001.1585[nucl-th]}.

\bibitem{Bozek}
Piotrz Bozek.
\newblock {\em arXiv:0911.2397[nucl-th]}.

\bibitem{Schafer2009}
Thomas Schafer and Derek Teaney.
\newblock {\em Rep. Progr. Phys.}, 72:126001, 2009.

\bibitem{Arsene2005}
I.~Arsene and et.al. (BRAHMS~Collaboration).
\newblock {\em Nucl. Phys. A}, 757:1, 2005.

\bibitem{Back2005}
B.B. Back and et.~al. (PHOBOS~Collaboration).
\newblock {\em Nucl. Phys. A}, 757:28, 2005.

\bibitem{Adcox2005}
K.~Adcox and et~al. (PHENIX~Collaboration).
\newblock {\em Nucl. Phys. A}, 757:184, 2005.

\bibitem{Csernai2006}
L.P. Csernai, J.I. Kapusta, and L.D. McLerran.
\newblock {\em Phys. Rev. Lett.}, 97:152303, 2006.

\bibitem{Gorenstein2008}
M.I. Gorenstein, M.~Hauer, and O.N. Moroz.
\newblock {\em Phys. Rev. C}, 77:024911, 2008.

\bibitem{Khan}
E.~Khan.
\newblock {\em arXiv:0905.3335[nucl-th]}.

\bibitem{Maeda}
K.~Maeda, G.~Baym, and T.~Hatsuda.
\newblock {\em arXiv:0904.4372[cond-mat.quant.gas]}.

\bibitem{Xu}
Renxin Xu.
\newblock {\em arXiv:0912.0349[astro-ph.HE]}.

\bibitem{Taylor}
E.~Taylor and M.~Randera.
\newblock {\em arXiv:1002.0869[cond-mat.quant-gas]}.

\bibitem{Koide}
T.~Koide, E.~Nakano, and T.~Kodama.
\newblock {\em arXiv:0901.3707[hep-ph]}.

\bibitem{Angilella}
G.G.N. Angilella and et~al.
\newblock {\em arXiv:0901.265[cond-mat.stat-mech]}.

\bibitem{Fisher1964}
I.Z. Fisher.
\newblock {\em Statistical Theory of Liquids}.
\newblock University of Chicago Press, Chicago, 1964.

\bibitem{Boer1948}
J.~De Boer.
\newblock {\em Physica}, 14:139, 1948.

\bibitem{Landau1980}
L.~D. Landau and E.~M. Lifshits.
\newblock {\em Course of Theoretical Physics}, volume~5.
\newblock Pergamon, Oxford, 1980.

\bibitem{Bulavin2008}
L.~A. Bulavin, L.~L. Jenkovszky, V.~K. Magas, V.~M. Sysoev, and K.~V.Cherevko.
\newblock In {\em Proceedings of the 2-nd International Conference on Current
  Problems in Nuclear Physics and Atomic Energy}, volume~1, page 201. Kyiv
  Institute for Nuclear Research of NASU, 2008.

\bibitem{Witten1984}
E.~Witten.
\newblock {\em Phys. Rev. D}, 30:272, 1984.

\bibitem{Chandra2000}
D.~Chandra and A.~Goyal.
\newblock {\em Phys. Rev. D}, 62:063505, 2000.

\bibitem{Brilenkov2014}
M.~Brilenkov, M.~Eingorn, L.~Jenkovszky, and A.~Zhuk.
\newblock {\em Eur. Phys. J. C}, 74:3011, 2014.

\bibitem{Jenkovszky1990}
L.~L. Jenkovszky, B.~Kampfer, and V.~M. Sysoev.
\newblock {\em Z. Phys. C}, 48:147, 1990.

\bibitem{Boyko1990a}
V.~G. Boyko, L.~L. Jenkovszky, B.~Kampfer, and V.~M. Sysoev.
\newblock {\em Astron. Nachrichten.}, 311:265--269, 1990.

\bibitem{Bulavin2010a}
L.~A. Bulavin, L.~L. Jenkovszky, S.~M. Troshin, and N.~E. Tyurin.
\newblock {\em EChAYa}, 41:924--927, 2010.

\end{thebibliography}


\vskip-2mm
\end{document}